**ORIGINAL ARTICLE**

# *HST* astrometry of the closest Brown Dwarfs – II. Improved parameters and constraints on a third body [†]

L. R. Bedin*[1]  |  J. Dietrich[2,3]  |  A. J. Burgasser[4]  |  D. Apai[2,5]  |  M. Libralato[1,6]  |  M. Griggio[1,7,8]  |  C. Fontanive[9]  |  D. Pourbaix[10]

[1] Istituto Nazionale di Astrofisica, Osservatorio Astronomico di Padova, Vicolo dell'Osservatorio 5, Padova, IT-35122, Italy

[2] Department of Astronomy and Steward Observatory, The University of Arizona, 933 N. Cherry Avenue, Tucson, AZ 85721, USA

[3] School of Earth and Space Exploration, Arizona State University, 781 Terrace Mall, Tempe, AZ 85287, USA

[4] Department of Astronomy & Astrophysics, University of California San Diego, La Jolla, CA 92093, USA

[5] Lunar and Planetary Laboratory, The University of Arizona, 1629 E. University Blvd., Tucson, AZ 85721, USA

[6] AURA for the European Space Agency (ESA), Space Telescope Science Institute, 3800 San Martin Drive, Baltimore, MD 21218, USA

[7] Università di Ferrara, Dipartimento di Fisica, Via Giuseppe Saragat 1, I-44122, Ferrara, Italy

[8] Space Telescope Science Institute, 3800 San Martin Drive, Baltimore, MD 21218, USA

[9] Trottier Institute for Research on Exoplanets, Université de Montréal, 1375, ave. Thérèse-Lavoie-Roux, Montréal, Québec, H3C 3J7, Canada

[10] Institut d'Astronomie et d'Astrophysique,Université Libre de Bruxelles (ULB),B-1050 Bruxelles,Belgium

**Correspondence**
*E-mails: luigi.bedin@inaf.it

Located at less than 2 pc away, Luhman 16 AB (WISE J104915.57-531906.1) is the closest pair of brown dwarfs and third closest "stellar" system to Earth. An exoplanet candidate in the Luhman 16 binary system was reported in 2017 based on a weak astrometric signature in the analysis of 12 HST epochs. An additional epoch collected in 2018 and re-analysis of the data with more advanced methods further increased the significance level of the candidate, consistent with a Neptune-mass exoplanet orbiting one of the Luhman 16 brown dwarf components. We report the joint analysis of these previous data together with two new astrometric *HST* epochs we obtained to confirm or disprove this astrometric signature.

Our new analysis rules out presence of a planet orbiting one component of the Luhman 16 AB system for masses $\mathcal{M} \gtrsim 1.5\,\mathrm{M_{\Psi}}$ (Neptune masses) and periods between 400 and 5000 day. However, the presence of third bodies with masses $\mathcal{M} \lesssim 3\,\mathrm{M_{\Psi}}$ and periods between 2 and 400 days (∼1.1 yrs) can not be excluded.

Our measurements make significant improvements to the characterization of this sub-stellar binary, including its mass-ratio 0.8305±0.0006, individual component masses 35.4±0.2 $\mathrm{M_{\jupiter}}$ and 29.4±0.2 $\mathrm{M_{\jupiter}}$ (Jupiter masses), and parallax distance 1.9960 pc ± 50 AU. Comparison of the masses and luminosities of Luhman 16 AB to several evolutionary models shows persistent discrepancies in the ages of the two components, but strengthens the case that this system is a member of the 510±95 Myr Oceanus Moving Group.

# 1 | INTRODUCTION

Astronomers using the *Wide-field Infrared Survey Explorer* (WISE; Wright et al. 2010) discovered many of the nearest ($d < 20$ pc) and coldest (T < 1000 K) brown dwarfs (BDs) in the Solar neighborhood (Kirkpatrick et al., 2011). The Luhman 16 AB (WISE J104915.57-531906.1) system is one such discovery, a binary brown dwarf in the Southern constellation Vela (Luhman, 2013). At a distance of only ∼2 pc, Luhman 16 AB are the closest known brown dwarfs to Earth, and the closest 'stellar' objects found since Barnard's Star was identified in 1916.

The primary (Luhman 16A) is of spectral type L8±1 and the secondary (Luhman 16B) is of spectral type T1±2, so the system straddles the L dwarf/T dwarf transition (Burgasser, Sheppard, & Luhman, 2013; Kniazev et al., 2013). Luhman 16 A and B orbit each other at a projected separation of approximately 3.5 AU with an orbital period of about 30 years (Garcia et al., 2017).

Its proximity and low combined mass make the Luhman 16 AB system exceptionally well-suited for astrometric reconnaissance. The *Hubble Space Telescope (HST)* multi-cycle general observers (GO) programs GO 13748 and GO 14330 (PI Bedin) were designed to achieve very high-precision astrometry, and measured the binary brown dwarfs' positions over 13 epochs spanning 4 yr. The program's goal was to refine measurement of the system's orbital and physical, parameters and to search for the astrometric signatures of a third body. Bedin et al. (2017, hereafter Paper I) presented data from 12 out of the 13 planned *HST* epochs, sampling positions over a temporal baseline of 2.1 years between 2014.64 and 2016.76. These data significantly improved the astrometric parameters for Luhman 16 AB, and excluded the presence of a third body down to sub-Neptune masses (∼10 $M_\oplus$) with periods between 1-2 years, and ruled out the presence of a giant planet candidate proposed by Boffin et al. (2014) based on ground-based measurements, also excluded by Sahlmann and Lazorenko (2015).

However, Paper I did report the detection of astrometric residuals compatible with a *sub-Neptune* planet with an orbital period longer than two years, which were interpreted as likely due to residual systematic errors. The final epoch of program GO-13748+14330 was collected in August 2018, extending the temporal baseline to about 4 years. These observations boosted the sensitivity for long-term variations. We reduced and re-analyzed all 13 epochs from GO-13748+14330 in a self-consistent manner, using more advanced techniques than those deployed in Paper I. In particular, we improved the astrometry by linking measurements to the absolute reference system of Gaia-DR3 (Gaia Collaboration et al., 2023), based on procedures similar to those described in Bedin and Fontanive (2018, 2020). With the extended baseline and more sophisticated analysis, the residual from the A-B orbital fit detected by Paper I was found to be even more pronounced. A search for periodic signals with the Generalized Lomb-Scargle (GLS) period-searching algorithm (Zechmeister & Kürster, 2009) identified a periodic signal with a formal false alarm probability of $2\times10^{-8}$, a marginally significant amplitude of 0.7±0.3 mas, and a period of ∼2.3 yrs. By adopting masses of 34±1 $M_{\jupiter}$ and 28±1 $M_{\jupiter}$ for Luhman 16 A and B, respectively (Garcia et al., 2017), the best-fit astrometric signal corresponds to an exoplanet with a mass just below that of Neptune. Support for the genuine nature of the signal was also provided by the fact that there is no known systematic error or telescope/instrument state parameter that would vary with a 2.3 yr periodicity. On the other hand, a tertiary with this long of an orbital period would not be long-term stable as a hierarchical triple (Eggleton & Kiseleva, 1995).

The possibility of a sub-Neptune planet orbiting within the Luhman 16 AB system would be an exciting and scientifically important discovery for stellar, brown dwarf, and exoplanetary research, given the system's proximity and our ability to measure precise host masses. However, an extension of the data coverage over at least two 2.3 yr periods was needed to confirm or disprove the presence of this candidate. We therefore collected two additional epochs of *HST* imaging in program GO-15884 (PI Bedin) with the same observing strategy as the previous programs timed to maximize the putative astrometric signal of this exoplanet candidate.

In this paper, we present the new *HST* observations, and conduct a joint analysis of all astrometric measurements of the Luhman 16 AB system in combined dataset. In Section 2, we discuss the new observations, In Section 3, we describe the data reduction and astrometric measurements, conducted using an updated approach to that described in Paper I that makes use of *Gaia* DR3 (Gaia Collaboration et al., 2023). In Section 4, we describe measurement of the barycentre motion of the system, including improved measurement of the system's parallax and proper motion, and inference of the system mass ratio. In Section 5, we determine improved orbit parameters of the Luhman 16 AB pair, including total and component masses, based on data that now span 180° of projected true anomaly phase coverage of the orbit around periastron. In Section 6, we evaluate evidence for a third component in the system using a perturbative Markov chain Monte Carlo *(MCMC)* approach, ruling out planets above 3 Neptune masses. In Section 7, we place these findings in the context of our understanding of the Luhman 16 AB system. In Section 8, we summarize our results.

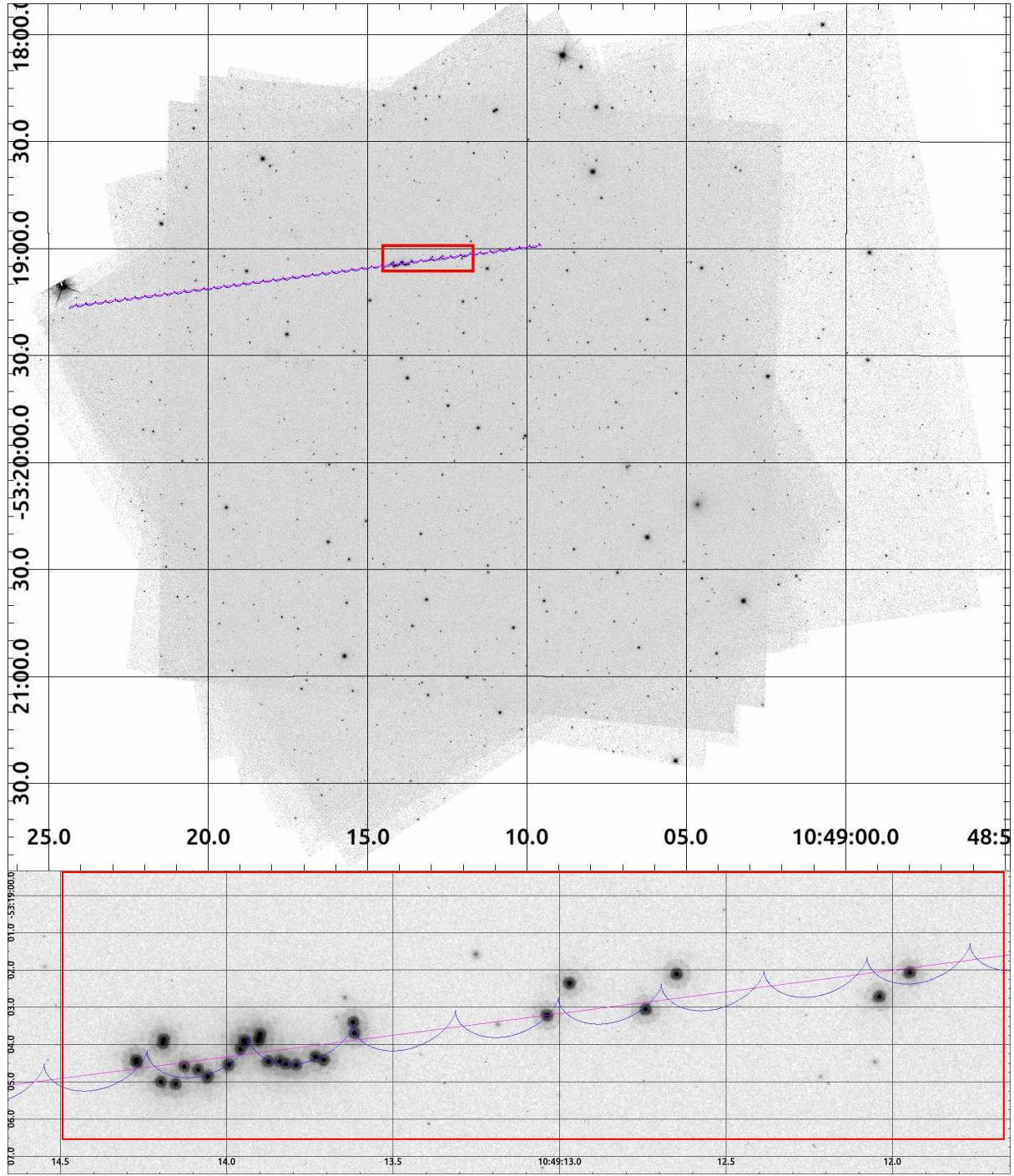

**FIGURE 1** *(Top:)* The 4.7′×4′ region surrounding Luhman 16 AB, as monitored by *HST*. The image is the sum of the stacks in WFC3/UVIS/F814W obtained for each of the 15 individual visits, each made from two images. *(Bottom:)* Zoom-in of the 25″×7″ rectangular region indicated in red in the top panel, showing the complete pattern in the sky of Luhman 16 A and B during the period monitored by our *HST* observations. Our astrometric solution for the Luhman AB barycenter, with (and without) parallax is indicated in blue (magenta).

## 2 | NEW OBSERVATIONS

In this work, as in its precursor study (Paper I), we used only images obtained in staring-mode; i.e., keeping the *HST*

telescope as fixed on the field, which is the mode for the vast majority of *HST* imaging. The full dataset also includes trailed-images, in which the telescope slews in a controlled manner while exposing, enabling very high precision astrometry (Riess, Casertano, Anderson, MacKenty, & Filippenko, 2014). The treatment of these data is more complicated and is deferred to a future work (Libralato et al. 2024, in preparation). We point the interested readers to Paper I for an extensive description of the observing strategy.

Here we report three new *HST* epochs, one from GO-14330 and two from GO-15884, obtained with *Wide Field Camera 3 (WFC3)*. In each epoch, two 60 s images were collected with the *Ultra-Violet and Visual (UVIS) channel* in filter F814W, and one 340 s image in filter F606W, for a total of nine new images. Our analysis is based on the combined set of 45 images collected in 15 epochs, ranging from 2014.7 to 2022.1 and hence spanning a temporal baseline of $\Delta t \sim 7.4$ years. Figure 1 shows the full WFC3/UVIS field of view (FOV) as a stack of the 15 epochs in filter F814W, as well as a zoom-in of the region around the trajectory of Luhman 16 AB.

## 3 | DATA REDUCTION AND MEASUREMENTS

We follow the same procedure as in Paper I, which we summarise here with details on improvements made.

To reduce potential systematic errors in astrometry related to imperfections in the WFC3/UVIS detector's charge transfer efficiency (CTE), we adopted active and passive mitigations. We used the `_flc` images which apply the pixel-based CTE correction algorithms similar to those developed by Anderson and Bedin (2010), and as part of the observations injected artificial charge via post-flashing (of about $12\,e^-$) to fill most of the charge traps. Both strategies do not work perfectly, and show residuals at the $\sim 0.5$ mas-level in raw positions that can be tracked down and corrected as explained in Paper I.

Fluxes and positions were extracted from each `_flc` image using the publicly available[1] software `hst1pass` (Anderson, 2022), and state-of-the art geometric distortion and point spread function (PSF) libraries from Anderson (2022). The PSFs are given as spatially-varying models in 7×8 arrays, and were tailored for each individual image using the prescription of Anderson and Bedin (2017) to account for small focus variations across the whole FOV. Along with flux and position, `hst1pass` provides a diagnostic of the quality of the PSF-fitting, called `Q`, which measures how well a star is represented by the perturbed PSF local model. The `Q` parameter is close to zero for the best-fit point sources, and deviates for cosmic rays, detector cosmetic defects, diffraction artifacts, galaxies, and blends. The Luhman 16 components do have larger `Q` values ($\approx 0.07$) compared to other, bluer-color point sources in the images of comparable brightness ($\approx 0.02$), noticeable in the PSF-subtracted images. These deviations could be due to color dependencies in the PSF models and/or chromatic effects across the wide F606W and F814W filters that lead to systematic offsets for the target sources, which have significantly different spectral shapes than reference field stars (see also Bellini, Anderson, and Bedin 2011). However, position residuals for Luhman 16 AB were comparable to the expected random error for this camera (i.e. $\sim 320\,\mu$as). Furthermore, in the computation of the relative orbit, the two components have similar red-optical spectra (Kniazev et al., 2013), and should be affected the same way, canceling such biases.

The extracted raw positions of sources in individual frames ($x^{\mathrm{raw}}, y^{\mathrm{raw}}$) were then corrected for geometric distortion with the best available corrections (Bellini et al., 2011; Bellini & Bedin, 2009). The main improvement in this study compared to Paper I, is the way the reference frame was derived. Here, we followed the procedures outlined in latest version of the method developed by Bedin and Fontanive (2018, 2020), which links directly to the absolute International Celestial Reference System (ICRS). The method uses a combination of *HST* and *Gaia* to extend the astrometric accuracies of the latter to sources like brown dwarfs that are too faint for *Gaia* measurement. The method was tested and successfully demonstrated to reach sub-mas accuracies (Fontanive, Bedin, & Bardalez Gagliuffi, 2021).

We adopted 8 January 2016 (2016.02) as the reference epoch for the observational plane of positions, which is closest to the reference epoch of the *Gaia* DR3 catalog (2016.0). We found many well-measured sources in the 45 individual *HST* images employed in this work in the *Gaia* DR3 catalog, with between 72 and 164 (an average of 133±25 and a median of 147) sources in the WFC3/UVIS FOVs. Those *Gaia* sources with full 5-parameter astrometric solutions can be placed into the ICRS frame at each individual *HST* epoch. Linear transformations then link the coordinates within each individual *HST* image to the *Gaia* ICRS. This procedure involves going back and forth between projected and tangential planes, using methods and equations detailed in Section 3 of Bedin and Fontanive (2018). These transformations allowed us to place all observations into a common reference frame. We were then able to create stacked images for all the observations, such as the one shown in Figure 1 for the F814W filter.

In the next two sections (Sect. 4 and 5) we describe the two-step procedure that allowed us to derive the astrometric parameters of the system as a whole, and then the orbital parameters of the binary.

---

[1] https://www.stsci.edu/~jayander/HST1PASS/

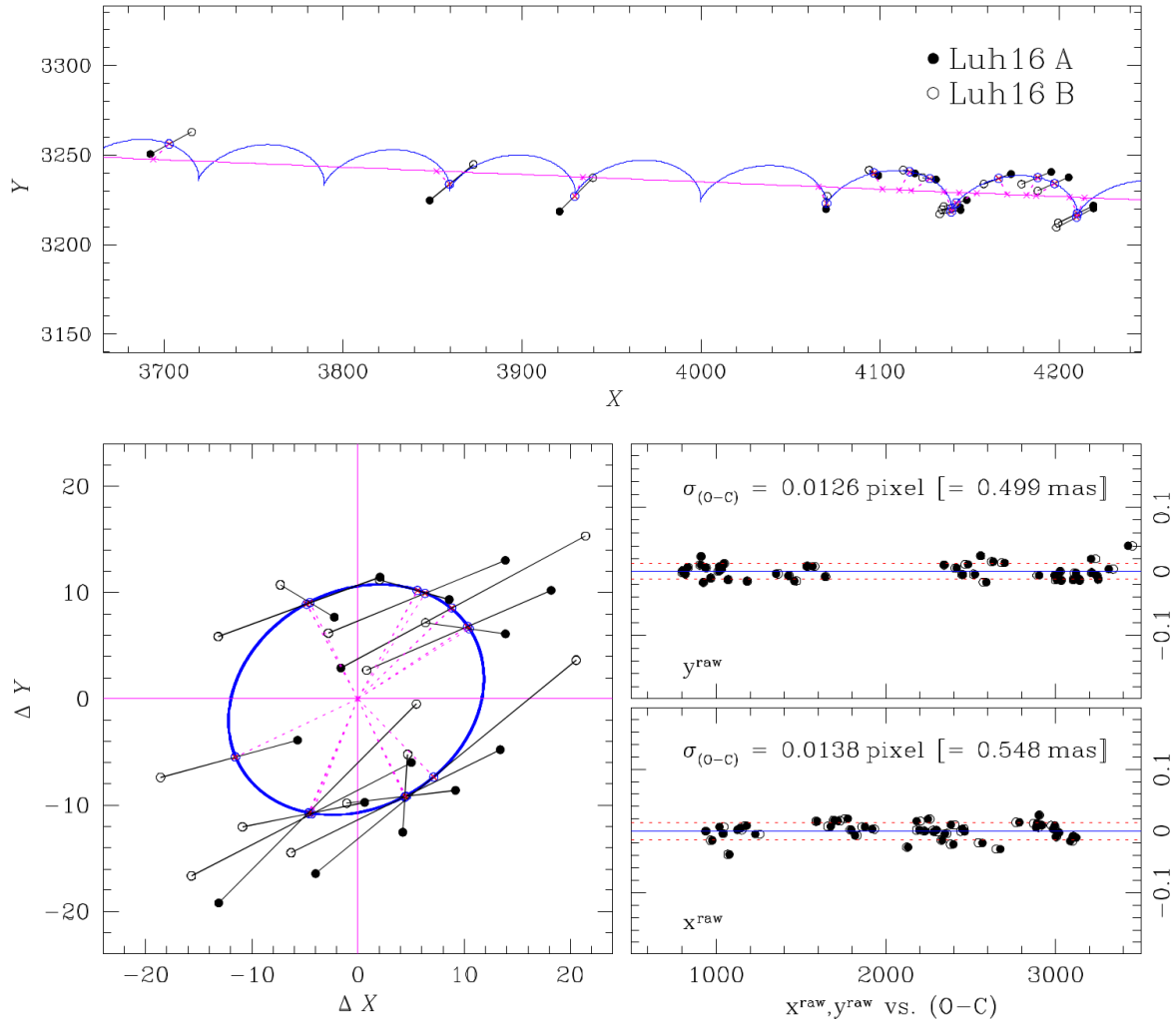

**FIGURE 2** The top panel shows the complete series of observed positions for Luhman 16 A (filled circles) and Luhman 16 B (empty circles) in the coordinate system of our master frame. At any given epoch, a black line joins A and B, and along it the position of the baricenter G, estimated as explained in the text, is marked with small blue circles. A blue curve indicates the astrometric solution for Luhman 16 AB's baricenter (positions, proper motions, parallax), noting that the mass ratio, $q \equiv \mathcal{M}_{\rm B}/\mathcal{M}_{\rm A}$, is part of the solution. The line in magenta indicates the same solution with the system at an infinite distance, and crosses in magenta on this line indicate the individual epochs along this path. These locations are connected with the observed baricentric positions (dashed lines in magenta). The orientation in this plot is in pixels of the master frame, rather than R.A. and Dec. as in Fig. 1 . The bottom-left panel shows the same plot after removing the G proper motion to highlight the parallax contribution. Panels on bottom right show residuals as function of the pixel raw coordinates after the astrometric model for G was subtracted.

## 4 | DETERMINING THE ASTROMETRIC PARAMETERS

To derive the astrometric parameters, we proceeded in the manner described in Paper I. Here, we provide a brief overview of the relevant steps, and details for significant improvements.

Figure 1 illustrate the targets' projected motions for the 15 epochs collected between 22 August 2014 and 21 January 2022. The combination of non-uniform temporal sampling, parallax, proper and orbital motions makes it unclear where individual epochs are. A clearer view of the sources at the various epochs is provided by Fig. 2 , where positions in pixel coordinates of the reference epoch of the primary (A) and the secondary (B) are connected through the baricenter of the system (hereafter, "G"), which is not a direct observable.

At any given *HST* 1-orbit visit, there are actually three individual exposures taken less than an hour apart, which we treat as individual epochs, even if they are not visually distinguishable at this scale. Therefore, we used 45×2D data points to derive the six astrometric absolute parameters for the baricenter of the Luhman 16 AB system: positions ($\alpha,\delta$), proper motions ($\mu_{\alpha^*},\mu_\delta$), annual parallax ($\varpi$), and the mass-ratio ($q \equiv \mathcal{M}_B/\mathcal{M}_A$). The last parameter is treated as an astrometric parameter as it is inferred from the relative A-B positions from the barycenter. We have neglected astrometric radial variation, which induces time variation of the annual parallax and proper motions, as current spectroscopic estimates of the radial velocities are around 20 km/s (Crossfield et al., 2014; Kniazev et al., 2013; Lodieu et al., 2015), which would induce biases of only few $\mu$as/yr (see Bedin & Fontanive, 2020, for the calculation in the most extreme case, i.e., *Barnard's Star*).

**TABLE 1** Astrometric parameters of the baricenter (*G*) and mass-ratio of Luhman 16 AB in ICRS.

| | | |
|---|---|---|
| $q = \mathcal{M}_B/\mathcal{M}_A$ | 0.8305 | $^{+0.0006}_{-0.0006}$ |
| $\alpha_{\rm ICRS}$ [degrees] | 162.32821578 | $^{+0.9}_{-1.6}$ mas |
| $\delta_{\rm ICRS}$ [degrees] | −53.31941054 | $^{+0.8}_{-0.5}$ mas |
| $\alpha_{\rm ICRS}$ [hh mm ss] | $10^h49^m18.^s771$ | |
| $\delta_{\rm ICRS}$ [° ′ ″] | −53°19′09″.8779 | |
| $\mu_{\alpha^*_{\rm ICRS}}$ [mas yr$^{-1}$] | −2768.511 | $^{+0.056}_{-0.030}$ |
| $\mu_{\delta_{\rm ICRS}}$ [mas yr$^{-1}$] | 358.472 | $^{+0.027}_{-0.047}$ |
| $\varpi$ [mas] | 500.993 | $^{+0.059}_{-0.048}\pm0.050^\ddagger$ |
| distance [pc] | 1.996036 | $^{+0.00019}_{-0.00024}\pm 0.0002^\ddagger$ |
| distance [AU] | 411,712 | $^{+39}_{-48}\pm 41^\ddagger$ |

$^\ddagger$ Intrinsic accuracy of *Gaia* DR3 (Lindegren et al., 2021).
Note $\mu_{\alpha^*}$ is the compact symbology for $\mu_{\alpha\cos\delta}$.

Predicted positions of the baricenter were derived from the Naval Observatory Vector Astrometry Software (hereafter NOVAS; version F3.1, Kaplan, Bartlett, Monet, Bangert, and Puatua 2011) provided by the U.S. Naval Observatory, which accounts for many subtle effects of Solar System dynamics. We used a Levenberg–Marquardt algorithm (the FORTRAN version `lmdif` available under Moré, Garbow, and Hillstrom 1980) to find the minimization of the six astrometric parameters: ($\alpha,\delta,\mu_{\alpha^*},\mu_\delta,\varpi,q$). In Figure 3 we show the density profiles obtained by propagating Gaussian error distributions for the observed positions through the parameters derivation (Monte-Carlo simulations, with about 1 000 realisations). Our final astrometric solution is illustrated in Fig. 2 and given in Table 1 . We conservatively included an intrinsic error of 50 $\mu$as to account for the local absolute accuracy of *Gaia* EDR3/DR3 parallaxes (Lindegren et al., 2021).

Figure 2 shows average residuals of ∼0.5 mas, slightly larger than expected for stars of this brightness for WFC3/UVIS, 0.008 pixels or 320 $\mu$as (Bellini et al., 2011). For relative measurements to determine G, the expected uncertainties are $\sqrt{2}$ larger, 0.011 pixels, or 450 $\mu$as. Again, we attribute these residuals to color-dependent effects between our intrinsically red targets that are distinct from the reference field stars, of particular importance for broad-band filters. To quantify this effect, we examined the positions of *Gaia* DR3 sources in the *HST* images with *Gaia* color range $0 < G_{\rm BP} - G_{\rm RP} < 2$, but found no detectable trend. However, three reference field stars with $G_{\rm BP} - G_{\rm RP} > 2$ do show a systematic change in *HST* positions as large as 0.5 mas. Given the large uncertainties in the parallaxes and photometry of these three sources, it is not possible to assess the significance of these potential systematic residuals. We therefore assert that our final absolute positions do not have color-dependent systematic errors significantly worse than about 0.5 mas.

## 5 | DETERMINATION OF THE BINARY ORBITAL PARAMETERS

With the system's astrometric parameters in hand, we now move to the orbital solution. In Fig. 4 we show the relative positions of Luhman 16 B with respect to the primary component Luhman 16 A. Note that with the new three *HST* epochs added to the ones presented in Paper I, our observations now span almost 180° of the projected true anomaly despite the fact that the time interval covered by our observations (∼7 yrs) is less than a quarter of the estimated orbital period (∼30 years). This relatively large phase coverage is because our observations cover the periastron of a highly-eccentric ($e \sim 0.4$) orbit.

We include in Fig. 4 the individual data point from the digitalized ESO plate collected in 1984 as measured by Garcia et al. (2017) and independently remeasured by us for this work (see APPENDIX A:). While we did not use these data to constrain our orbital solution given it poor precision, both measures are nevertheless within ∼1.5 $\sigma$ of the solution. *Gaia* measurements for Luhman 16 are not usable for the present investigation (see APPENDIX B:).

To determine the seven orbital parameters of this visual binary — angular semimajor axis (*a*), eccentricity (*e*), period (*P*), inclination (*i*), longitude of ascending node ($\Omega$), argument of periapsis ($\omega$), and epoch of periapsis ($T_0$) — we followed the trial-and-error approach described by Pourbaix (1994). We employed the software tools `union` and `epilogue` (Pourbaix, 1998b), which are described in detail with applications in Pourbaix (1998a, 1998b, 2000). Briefly, `union` and `epilogue` are used to simultaneously adjust the observations of double-lined spectroscopic and visual binaries, as follows:

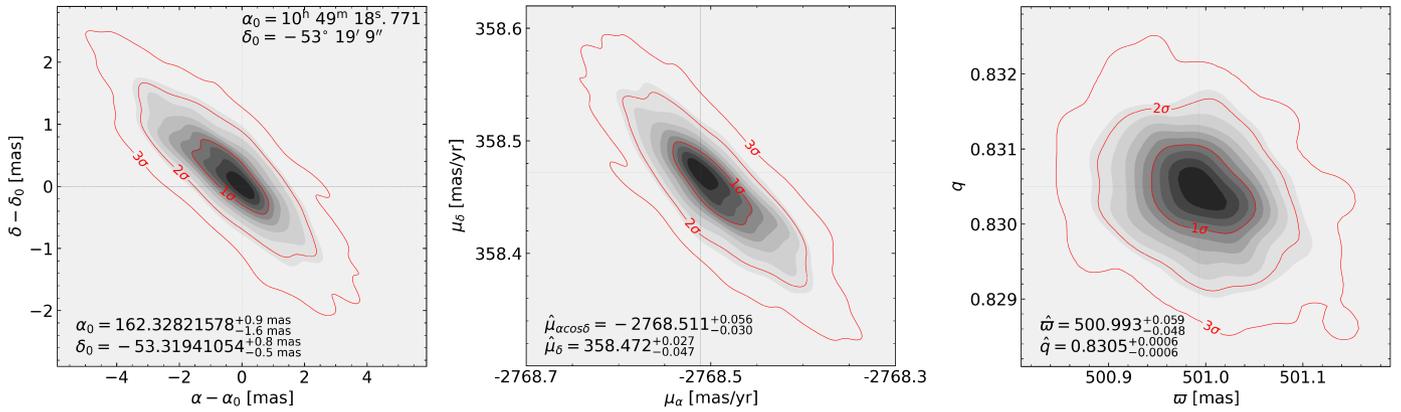

**FIGURE 3** Estimates for six barycenter astrometric parameters: (left) barycenter ICRS position at epoch 2000.0 with no parallax, $(\alpha_0, \delta_0)$, (center) proper motion components $(\mu_{\alpha\cos\delta}, \mu_\delta)$, and (right) parallax ($\varpi$) and mass ratio (q). Two-dimensional error confidence regions of $1\sigma$, $2\sigma$, and $3\sigma$ are indicated.

- `union` undertakes a global optimisation of the orbital solution followed by a preliminary local search. In order to increase the chance of finding the optimal solution, the user supplies constraints on parameters such as the orbital period.

- `epilogue` processes the results of `union` and generates statistical estimates of fit parameters, quality of fit, and ephemerides for the solution.

The resulting orbital solution for the Luhman 16 AB system is given in Table 2 and visualized in Fig. 4 . The figure highlights the enormous range of motion probed by this dataset, spanning approximately 180° of orbital phase and over 2" of relative motion. To estimate uncertainties, we again propagated position uncertainties using Gaussian distributions, generated 1 000 independent realisations of the data, and re-derived astrometric and orbital parameters. In Fig. 5 we show the confidence level of the orbital parameters obtained for these realisations.

For clearness, the final orbital parameters derived in this work are those in column 'data-fit' of Table 2 with the uncertainties as derived from MCs simulations in last column. In Sect. 6 we will derive parameters *also* with different tools, however those values should be regarded as an independent check.

In the bottom- and left-panel of Fig. 4 we show residuals in R.A. and Dec. of the astrometry relative to the orbital solution. These residuals are generally smaller than 1 mas, but become as large as 2 mas and are systematically larger at the ends of the orbit sampled by our observations. The insets on the top-left of the main panel show the R.A. and Dec. residuals as a function of the epochs expressed in Julian years.

These residuals also show systematic trends with epoch of observation, being largest at the start (R.A.) and end (Dec.) of our data set, but with no obvious periodicity within the probed temporal baseline. As the orbit derivation is a relative astrometric determination, these residuals are not likely to be driven by color effects on the PSF, as both brown dwarfs have similar spectral energy distributions in the *HST* filters deployed, and hence such biases should cancel out. If these signals were attributable to a third body, it would have to have an orbital period greater than $\sim$10 years, inconsistent with long-term stability within the host A-B binary (Eggleton & Kiseleva, 1995). We will further investigate this possibility in detail in the next section. It is more likely that these residuals arise from unidentified systematic measurement errors and/or incomplete coverage of the full orbit that ultimately affects the best-fit orbital solution. Continued long-term monitoring of the system astrometry over a full orbit may resolve these residuals if a third body is not responsible.

In Fig. 6 we show the derived confidence levels for the total mass of the system computed from distance and the fit semi-major axis, as well as individual component masses inferred from the mass ratio from the barycentric motion. The latter values are highly correlated, but nevertheless result in the most precise measures (less than 1% uncertainty) of individual brown dwarf mass to date (cf. Dupuy and Liu 2017).

## 6 | RESIDUALS *VS.* 3$^{\text{RD}}$ BODY

While continued residuals in the combined astrometric and orbital solution may be attributed to uncorrected systematics, it is necessary to place constraints on the potential perturbation of a third body in the system, most likely an exoplanet companion to one or both components. We implemented a *Markov chain Monte Carlo (MCMC)* method to sample the posterior distribution of the binary system parameters, in order to determine if adding a potential third body as a perturber of the orbit would explain the residuals from the previous fit. We set up two main orbital frameworks as hypotheses to test in the sampler. The first was simply a change in the orbital parameters of the binary itself, with no perturbing planetary companion, while the second added in a planetary mass object around one of the two bodies (both brown dwarfs were tested as a potential host for a planetary mass companion). This approach was followed because a three-body system cannot be solved analytically like a two-body system can, so numerical N-body integrations are required to determine if the residuals can be fit by a planetary companion perturbing the orbit of the binary.

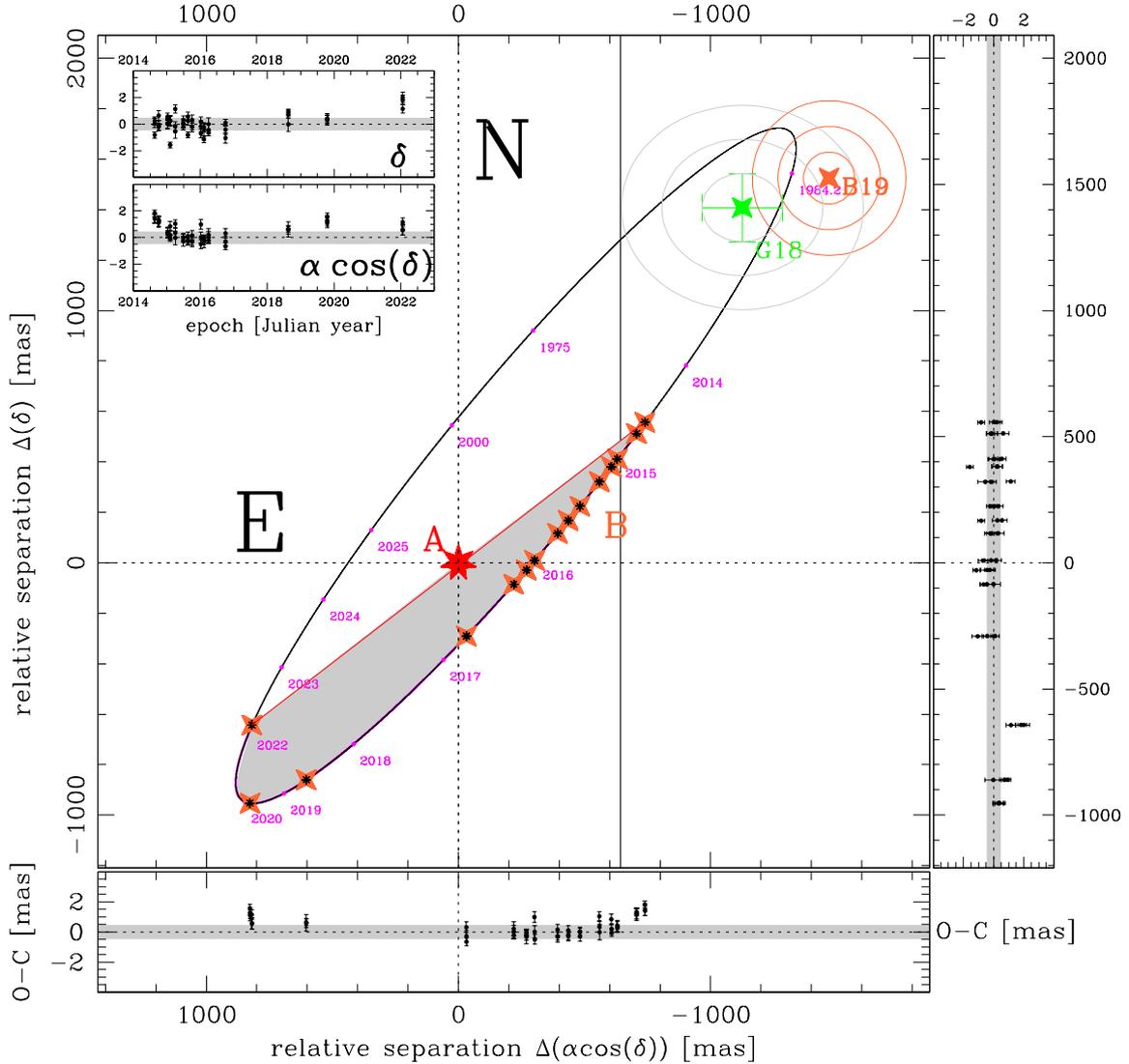

**FIGURE 4** The relative orbit of Luhman 16 B (four-pointed star symbols) around A (seven-pointed star symbol at origin). To better highlight the amount of curvature so far observed with *HST*, a red line connects the first and last observed relative positions, which span approximately 180° of orbital phase. For reference, a few epochs are indicated are along the orbit in magenta. The lower and right panels show the (O−C) residuals between the relative astrometry and our best-fit orbital model in mas. Grey bands signify the expected uncertainty for any individual relative positional measurement (i.e., $\sqrt{2}\times0.32$ mas). The insets in the large panel show the astrometric residuals in Right Ascension and declination as a function of observing epoch.

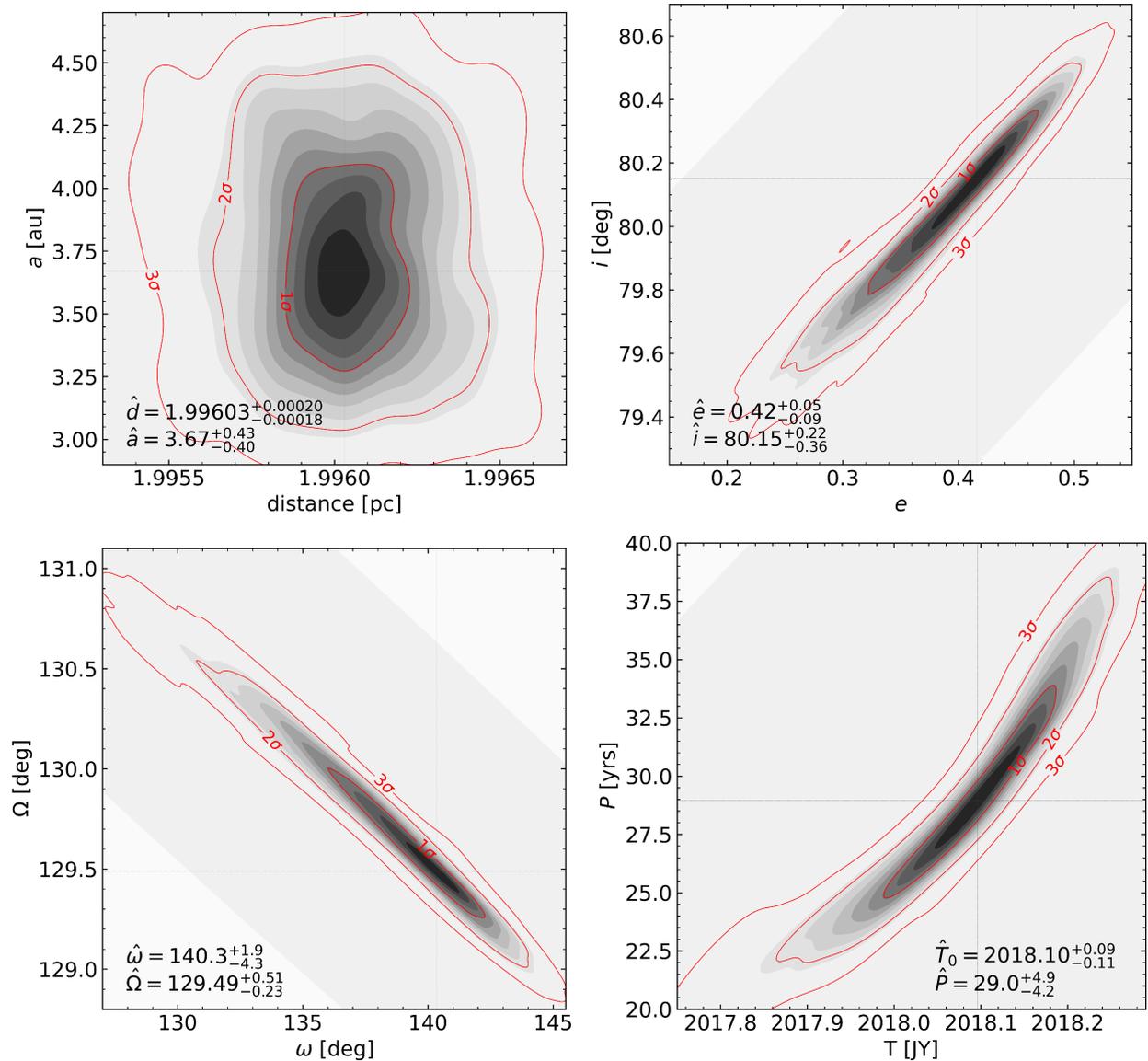

**FIGURE 5** Estimates for seven orbital parameters for the Luhman 16 AB binary, shown from left to right: system distance versus the projected semimajor axis $a$ (in AU), eccentricity $e$ versus inclination $i$, argument of periapsis $\omega$ versus longitude of ascending node $\Omega$, and epoch of periapsis $T_0$ versus period $P$ (in years). Two-dimensional error confidence regions of $1\sigma$, $2\sigma$, and $3\sigma$ are indicated.

We used constrained uniform priors on all of the parameters to avoid potential biases, and tested our fits with both fixed and free parameters for the barycenter and astrometric parameters (reference position, proper motion, parallax, and mass ratio). Our fits focused on the astrometry of the Luhman 16 AB components (and a possible third component) relative to the system barycenter. We assumed astrometric precisions of 0.5-1 mas in R.A. and Dec., and that measures were independent of each other and had uncertainties that were normally distributed, providing a functional form of a multivariate normal for the likelihood function to be maximized by the fitting process. For our free barycenter fit, we also verified that resulting barycenter positions were in agreement to within 100 $\mu$as with the values inferred from NOVAS, in this case using the Python version of the code (Barron et al., 2011) and its `astro_star` method.

We simulated the positions of the binary brown dwarfs and possible planetary companion with respect to the barycenter at each astrometric epoch using the N-body code `REBOUND` (Rein & Liu, 2012). In this scenario, in order to test if a 2-body or 3-body system fit the data better, we used a hybrid symplectic integrator for close planetary encounters called `MERCURIUS` (Rein et al., 2019), which is similar to the `MERCURY` integrator from Chambers (1999). `MERCURIUS` uses the fast symplectic Wisdom-Holman integrator `WHFast` (Rein & Tamayo, 2015) for a majority of the orbit. When a close encounter between two bodies occurs, it switches over to the high-accuracy non-symplectic Integrator with Adaptive Step-control, 15th order (`IAS15`; Rein and Spiegel 2015), which can provide orbital accuracy down to machine precision. By utilising this hybrid integrator, we are able to maximize the combination of speed and accuracy for the orbits and easily determine which configuration is the most likely in the system.

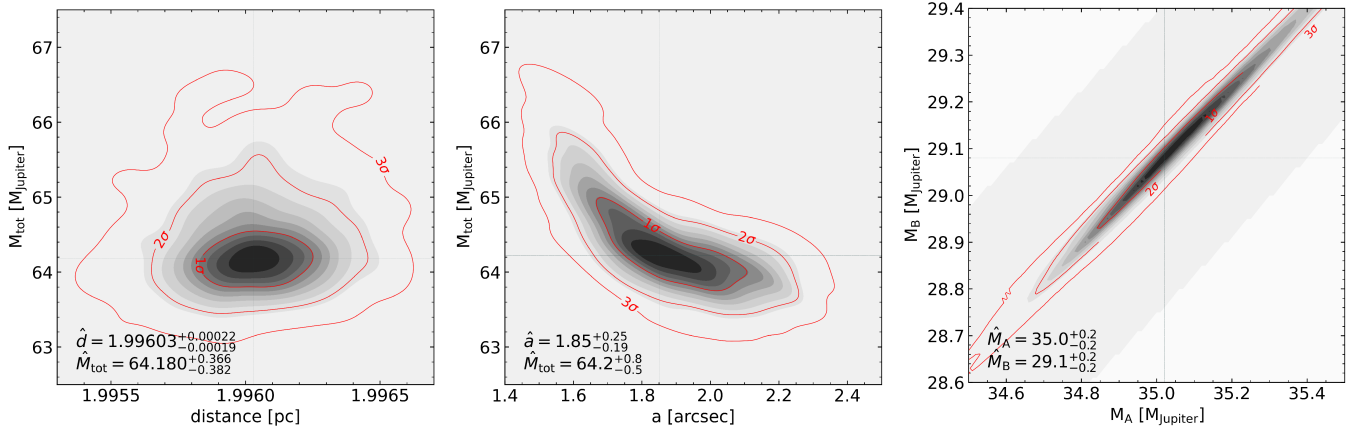

**FIGURE 6** Constraints on the masses of the brown dwarfs in the Luhman 16 AB binary: (left): total system mass $\mathcal{M}_{tot}$ in Jupiter mass units versus distance, (center): total system mass versus angular separation $a$, and (right): secondary mass $\mathcal{M}_B$ versus primary mass $\mathcal{M}_A$ in Jupiter mass units. Two-dimensional error confidence regions of $1\sigma$, $2\sigma$, and $3\sigma$ are indicated. The right panel shows a strong correlation in the derived masses for the two components.

**TABLE 2** The values of orbital parameters and masses of Luhman 16 AB, as derived from `union/epilogue` 'data-fit' to the actual data (shown in Fig. 4) with their formal uncertainties in parenthesis, and from MC simulations (as described in Sect.5) for the errors assessments (see text).

| parameter | data-fit | MCs$_{error-}^{error+}$ |
|---|---|---|
| $a$ [arcsec] | 1.764 ($\pm 0.003$) | $(1.85)_{-0.19}^{+0.25}$ |
| $(a\,[AU])^\star$ | 3.52 | $(3.67)_{-0.40}^{+0.43}$ |
| $i$ [deg] | 79.92 ($\pm 0.008$) | $(80.15)_{-0.36}^{+0.22}$ |
| $\omega$ [deg] | 136.67 ($\pm 0.09$) | $(140.3)_{-4.3}^{+1.9}$ |
| $\Omega$ [deg] | 130.02 ($\pm 0.01$) | $(129.49)_{-0.23}^{+0.51}$ |
| $e$ | 0.344 ($\pm 0.001$) | $(0.42)_{-0.09}^{+0.05}$ |
| $P$ [yr] | 26.55 ($\pm 0.08$) | $(29.0)_{-4.2}^{+4.9}$ |
| $T_o$ [Julian yr] | 2018.060 ($\pm 0.003$) | $(2018.10)_{-0.11}^{+0.09}$ |
| $\mathcal{M}_{tot}^\dagger$ [$M_\odot$] | 0.0619 | $(0.0613)_{-0.0004}^{+0.0003}$ |
| $\mathcal{M}_{tot}^\dagger$ [$M_{\jupiter}$] | 64.8 | $(64.2)_{-0.5}^{+0.8}$ |
| $\mathcal{M}_A^\ddagger$ [$M_{\jupiter}$] | 35.4 | $(35.0)_{-0.2}^{+0.2}$ |
| $\mathcal{M}_B^\ddagger$ [$M_{\jupiter}$] | 29.4 | $(29.1)_{-0.2}^{+0.2}$ |

$^\star$ Assuming a parallax of 500.993 mas.
$^\dagger$ Total mass, $\mathcal{M}_{tot} = \mathcal{M}_A + \mathcal{M}_B = a^3/P^2$.
$^\ddagger$ Assuming $q = 0.8305$, $\mathcal{M}_A = 1/(1+q)\mathcal{M}_{tot}$, and $\mathcal{M}_B = q/(1+q)\mathcal{M}_{tot}$.

We used the *MCMC* sampling framework `emcee` (Foreman-Mackey, Hogg, Lang, & Goodman, 2013) to optimize our fits. The only set of parameters to converge through the *MCMC* process was the 2-body binary with an adjustment to the sky position parameters. This is due to the extended temporal baseline with the new observations, which provides more accuracy and precision on the position of the binary in the sky. We found the best fit adjusts the positions of the stars at the J2000 epoch outside of the uncertainty from the original fits, but keeps the binary orbital parameters to within the original uncertainty.

For our three-body solution, we explored an exoplanet mass range of $25\,M_\oplus$ (or about $1.5\,M_{\jupiter}$) to $320\,M_\oplus$ ($\sim 1\,M_{\jupiter}$) and period range of 2-400 days. The outer limit of the period range comes from hierarchical triple system stability arguments (Eggleton & Kiseleva, 1995), which assuming a mass ratio $q_{C,AB} \equiv \mathcal{M}_C/\mathcal{M}_{A,B} \approx 0$ and $e_C \approx 0$ sets a maximum on the third body's semi-major axis of 0.35 AU, corresponding to a period of 1.1 yr. The lower limit of the mass range of $\sim 1.5\,M_{\jupiter}$ is derived from the semi-major axis upper limit and the expected astrometric signal accuracy of 0.5 mas for periods up to 400 days. For the third body's upper mass limit, one Jupiter mass ($1\,M_{\jupiter}$) was chosen, as it combines with the astrometric signal accuracy to give a lower limit for the orbital period of $\sim 2$ days. We provided a uniform distribution for each of the other parameters, with eccentricities between 0-1, inclinations between 0 and 180 degrees, longitude of ascending node and argument of periastron between 0 and 360 degrees, and epoch of periastron between 2000 and 2030. We found a range in the explored parameter space where the three-body fits did converge, but across most of the range it did not. Here we define "converging" vs. "not converging" by the Akaike Information Criterion (AIC). The AIC for the two-body fits and the three-body fits that did converge was ∼−4800, whereas the AIC for these three-body fits that did not converge was never lower than 6030. Thus, when the fits converge we cannot exclude a third body with those parameters.

We found that the three-body fits converged within the following ranges for the third body: mass between $25-50 M_\oplus$ (or ∼ $1.5-3 M_\Psi$), inclination between 60 and 90 degrees, eccentricity < 0.2, longitude of ascending node and argument of periastron between 90 and 180 degrees, and epoch of periastron between 2017 and 2023. In these cases, the brown dwarf binary orbital, mass, and astrometric parameters converged to the same values as the two-body fits. For completeness, we also tested the longer period range of 400-5000 days to attempt to exclude the original hypothesized third body at 2.3 years, although the astrometric signal caused by a $10 M_\oplus$ body at 2.3 years would only be ∼0.3 mas. We find that we could exclude all third bodies within this period range with masses of $25 M_\oplus - 1 M_{\lambda}$, as they did not converge and had a high AIC. Therefore, we conclude that if a third body is indeed orbiting one of the two components, it needs to have a mass below $50 M_\oplus$ orbiting within 0.35 AU (and periods shorter than 400 days) and with an orbital eccentricity less than 0.2.

Finally, we verified our barycenter plus orbit fit results (without a third body) by using a gradient optimization Trust Region Reflective fit (Branch, Coleman, & Li, 1999) with SCIPY's optimize.least_squares function instead of an MCMC sampler. Applying MCMC on the best-fit parameters from the gradient fit produced the parameter corner plot shown in Figure 7 and the values reported in Table 3 with their formal uncertainties. Several of the parameters show strong correlations, including the masses of the brown dwarfs, which makes sense given the constrained mass ratio between the two components; as well as the semi-major axis and the eccentricity, which is expected given our fit covers only half (in angular sweep) of the orbit. The semi-major axis and the orbital angles, including longitude of the ascending node, inclination angle and argument of periapsis, as well as and epoch of periastron are all also correlated, as all are tied together in order to fit the periastron passage. Most other parameters have weaker or no noticeable correlations between them.

We note the high degree of consistency for all astrometric parameters derived by this independent check with our final parameters derived derived in Sect. 4. However, there are significant differences in the derived orbital parameters compared to those obtained from the data-fit of the observations employing `union` and `epilogue.` in Sect. 5. These differences are along the strong correlation between the orbital parameters, and in any case within the confidence intervals defined by the MC simulations derived in Section 5 and shown in Fig. 5 .

**TABLE 3** Best-fit for astrometric- & orbital-parameters and masses for Luhman 16 AB with formal uncertainties, as derived from the *MCMC* simulations described in Sect. 6 and shown in Fig. 7 . The values in this table should be regarded as an independent check for those derived in Table 2 (see text).

| parameter | value | error+ | error− |
|---|---|---|---|
| $\alpha_{ICRS}$ [degrees] | 162.3282162 | +6.1 mas | −6.1 mas |
| $\delta_{ICRS}$ [degrees] | −53.3194099 | +6.5 mas | −6.1 mas |
| $\mu_{\alpha^*_{ICRS}}$ [mas yr$^{-1}$] | −2768.8551 | +0.0035 | −0.0036 |
| $\mu_{\delta_{ICRS}}$ [mas yr$^{-1}$] | 358.4482 | +0.0055 | −0.0055 |
| $\varpi$ [mas] | 501.023 | +0.077 | −0.079 |
| $a$ [AU] | 3.426 | +0.014 | −0.014 |
| $P$ [yr] | 25.508 | +0.158 | −0.157 |
| $i$ [deg] | 79.804 | +0.023 | −0.023 |
| $\omega$ [deg] | 135.13 | +0.22 | −0.23 |
| $\Omega$ [deg] | 130.143 | +0.027 | −0.027 |
| $e$ | 0.3262 | +0.030 | −0.030 |
| $T_o$ [Julian yr] | 2018.009 | +0.009 | −0.009 |
| $q = \mathcal{M}_B/\mathcal{M}_A$ | 0.8358 | +0.0022 | −0.0023 |
| $\mathcal{M}_{tot}$ [M$_{\lambda}$] | 64.742 | +0.087 | −0.089 |
| $\mathcal{M}_A$ [M$_{\lambda}$] | 35.266 | +0.067 | −0.068 |
| $\mathcal{M}_B$ [M$_{\lambda}$] | 29.476 | +0.056 | −0.057 |

## 7 | DISCUSSION

### 7.1 | Ruling out exoplanet companions in Luhman 16 AB

Our analysis firmly excludes the presence of a third body orbiting either component of the Luhman 16 AB binary with orbital periods greater than 400 days and masses greater than $25 M_\oplus$. However, we cannot rule out planetary companions with orbital periods between 2 and 400 days and masses less than $50 M_\oplus$. These constraints thus cannot rule out the existence of any habitable-zone terrestrial planets around either component ($a \approx 0.01$ AU, $P \approx 2$ days), as the expected astrometric signal for such a planet would be ∼1-10 $\mu$as. It is also conceivable that a third body of planetary-mass may exist in a circumbinary configuration (i.e., a planet orbiting both components; Doyle et al. 2011). Analysis of stable circumbinary orbits indicates that the minimum orbital period for such a system should be ~75 years (Holman & Wiegert, 1999), a period too long to cause the astrometric residuals in our sampled time interval. Nevertheless, continued monitoring over the coming decades may provide robust constraints for this scenario.

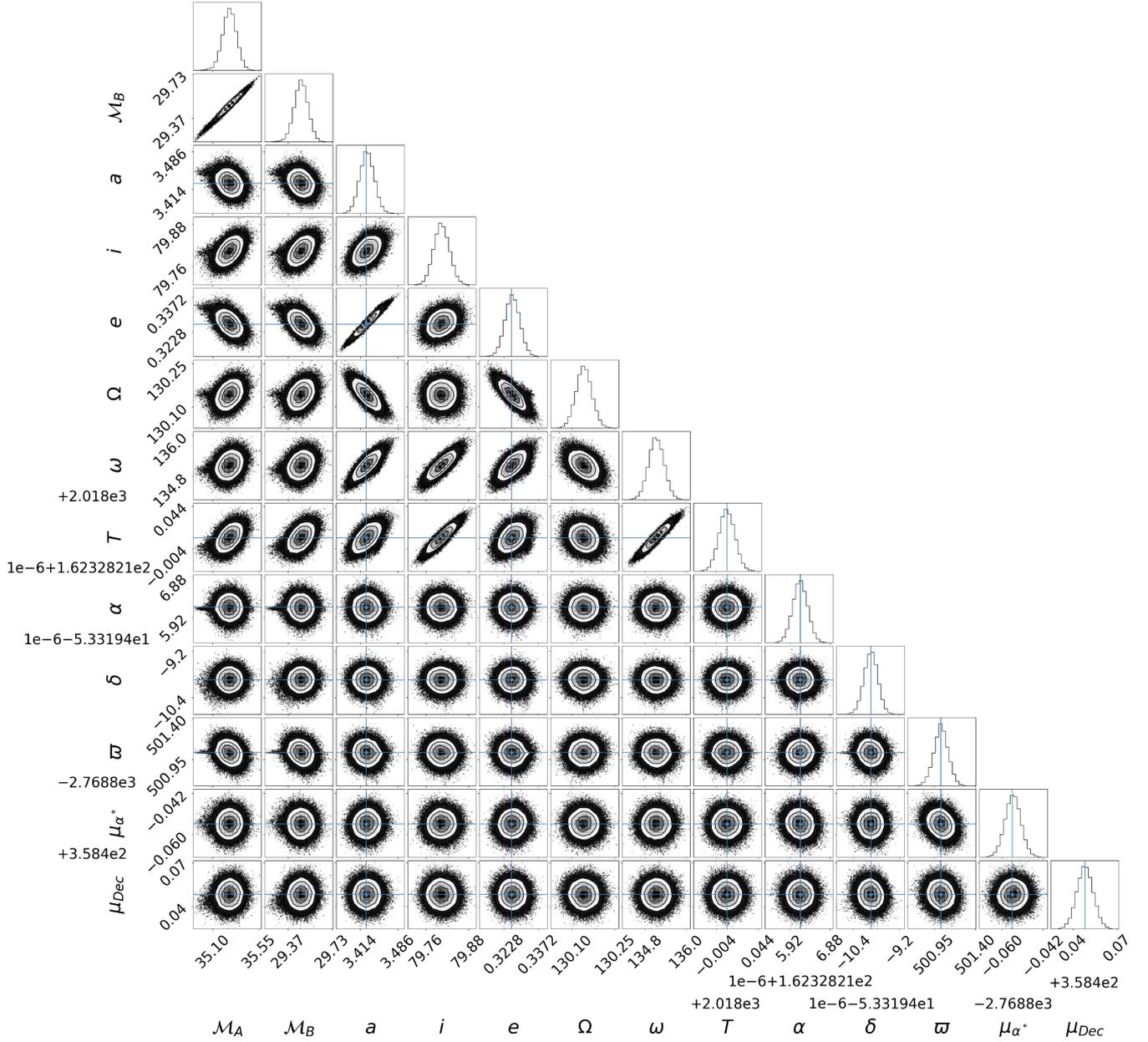

**FIGURE 7** The Corner plot (Foreman-Mackey, 2016) for the N-body integrations for the orbit of Luhman 16 AB based on a gradient optimisation fit to the 2-body orbital solution plus astrometric parameters, with an *MCMC* sampler to model the distributions. Parameters shown are primary mass ($m_1$ in $M_{\jupiter}$), secondary mass ($m_2$ in $M_{\jupiter}$), projected semimajor axis ($a$ in AU), inclination angle ($i$ in degrees), eccentricity ($e$), longitude of ascending node ($\Omega$ in degrees), argument of periapsis ($\omega$ in degrees), epoch of periastron passage ($T_0$ in years), system R.A. and Decl. at the reference epoch, system parallax in mas, and R.A. and Decl. in mas/yr. Diagonal plots display the marginalized distributions of each parameter, while interior plots show correlations. The contours and shadings correspond to the 2D Gaussian $0.5\sigma$-$1\sigma$-$1.5\sigma$-$2\sigma$ values, which cover ~11.8%, 39.3%, 67.5% and 86.4% of the 2D density, respectively.

## 7.2 | Refined System Parameters and Coevality Test of Evolutionary Models

Despite ruling out the presence of an exoplanet in the Luhman 16 AB system, our measurements nevertheless provide valuable insight into the nature of this nearest substellar system. In particular, we have improved measurement of its parallax (inaccessible from *Gaia*) to one part in 10 000, its mass ratio to one part in 1 300, and its total system mass to one part in 100 (Table 1 and 2 ). The last two values provide individual component masses with 1% precision, a quarter of the error on prior measurements for Luhman 16 AB by Garcia et al. (2017), and make these the most tightly constrained masses so far reported for resolved brown dwarfs. The system and component mass measures will improve as future astrometric measurements constrain the full orbit and reduce uncertainties on semimajor axis and period.

These accurately measured parameters provide the opportunity to test evolutionary models via a coevality test (cf. Liu, Dupuy, and Leggett 2010). Assuming the components of Luhman 16 AB were formed together, evolutionary models should predict the same age for both sources given their masses and luminosities. We examined seven sets of solar-metallicity evolutionary models from Baraffe, Chabrier, Allard, and Hauschildt (2001); Baraffe et al. (2003); Burrows et al. (2001); Chabrier et al. (2000); Phillips et al. (2020); Saumon and Marley (2008); and Marley et al. (2021). We interpolated the model grids for the component masses and uncertainties listed Table 2 as a function of age using tools in the `evolve` package of the SpeX Prism Libraries Analysis Toolkit (SPLAT; Burgasser and Splat Development Team 2017). We used separate interpolations for different cloud treatments in the models reported in Baraffe et al. (2001, cond versus dusty) and in Saumon and Marley (2008, no clouds, $f_{sed} = 2$, and hybrid), and different assumptions for chemical equilibrium versus non-equilibrium (which showed no effect on evolution) and choice of equation of state (Chabrier et al., 2019) in the models reported in Phillips et al. (2020). We compared the model luminosities to measured component luminosities reported by Faherty et al. (2014), which are generally more robust than effective temperature estimates which depend on spectral model fits and/or assumed radii.

Figure 8 displays a subset of these model comparisons, where the regions of $1\sigma$ overlap between the evolutionary models and measured luminosities are outlined in black. It is clear that the luminosity measurements are the dominant source of uncertainty in the inferred ages, which fall in the 400–800 Myr range for nearly all of the evolutionary models evaluated. This range is slightly younger than the 600–800 Myr range cited in Garcia et al. (2017), and is more in line with the estimated age of the Oceanus Moving Group (510±95 Myr), in which Gagné et al. (2023) argue Luhman 16 AB is a member system. Similar to Garcia et al. (2017), we find the inferred ages of the two components to be slightly mismatched, with Luhman 16 B being consistently younger than Luhman 16 A for every model, with no or minimal overlap in the $1\sigma$ regions. This mismatch may be due to errors in the evolutionary models or inaccuracies in the luminosities. An additional possibility is that one-dimensional evolutionary models may not capture correctly the cooling of brown dwarfs with heterogeneous cloud cover. Indeed, Luhman 16 A and B are known to have highly heterogeneous cloud cover and with varying extent of rotational asymmetry (e.g., Apai, Nardiello, & Bedin, 2021; Gillon et al., 2013; Karalidi, Apai, Marley, & Buenzli, 2016). The different cloud covering fraction between components A and B may introduce an apparent age mismatch due to different cooling efficiency.

Overall, there is minimal variation between the predictions of these models, with the single exception of the Saumon and Marley (2008) hybrid model which predicts significantly older ages for both components (∼650 Myr for Luhman 16 B and ∼900 Myr for Luhman 16 A). If the system is a member of Oceanus, our analysis rules out this specific model set based on lack of coevality and inaccurate ages. More rigorous constraints on the other evolutionary models will require improvements in the component luminosities, potentially through mid-infrared measurements from *JWST*.

## 8 | CONCLUSIONS

We have presented astrometric measurements and analysis from *HST* observations of the Luhman 16 AB system covering 45 images over 15 epochs spanning over 7.4 years. The key results of our study are as follows:

1. The astrometric parameters for Luhman 16 AB system derived in the absolute ICRS system provide the tightest constraints to date on its mass-ratio (better than 1 part in 1 300) and absolute parallax (better than 1 part in 10 000), corresponding to a distance of 1.9660 pc ± 50 AU.

2. Our measurement of the orbital motion, covering 180° of phase coverage, further provides the the most precise constraints on the orbital motion and component masses of any resolved brown dwarf system to date. In particularly, we constrain the individual dynamical masses down to 1%. Observations after 2020 enable us to observe the inversion of the orbital motion in the plane of the sky (cfr. Fig. 4 ), and effectively remove the so-called "ambiguity of quadrant" when determining the ellipse from a short observed arc of visual binaries (e.g., Heintz, 1978).

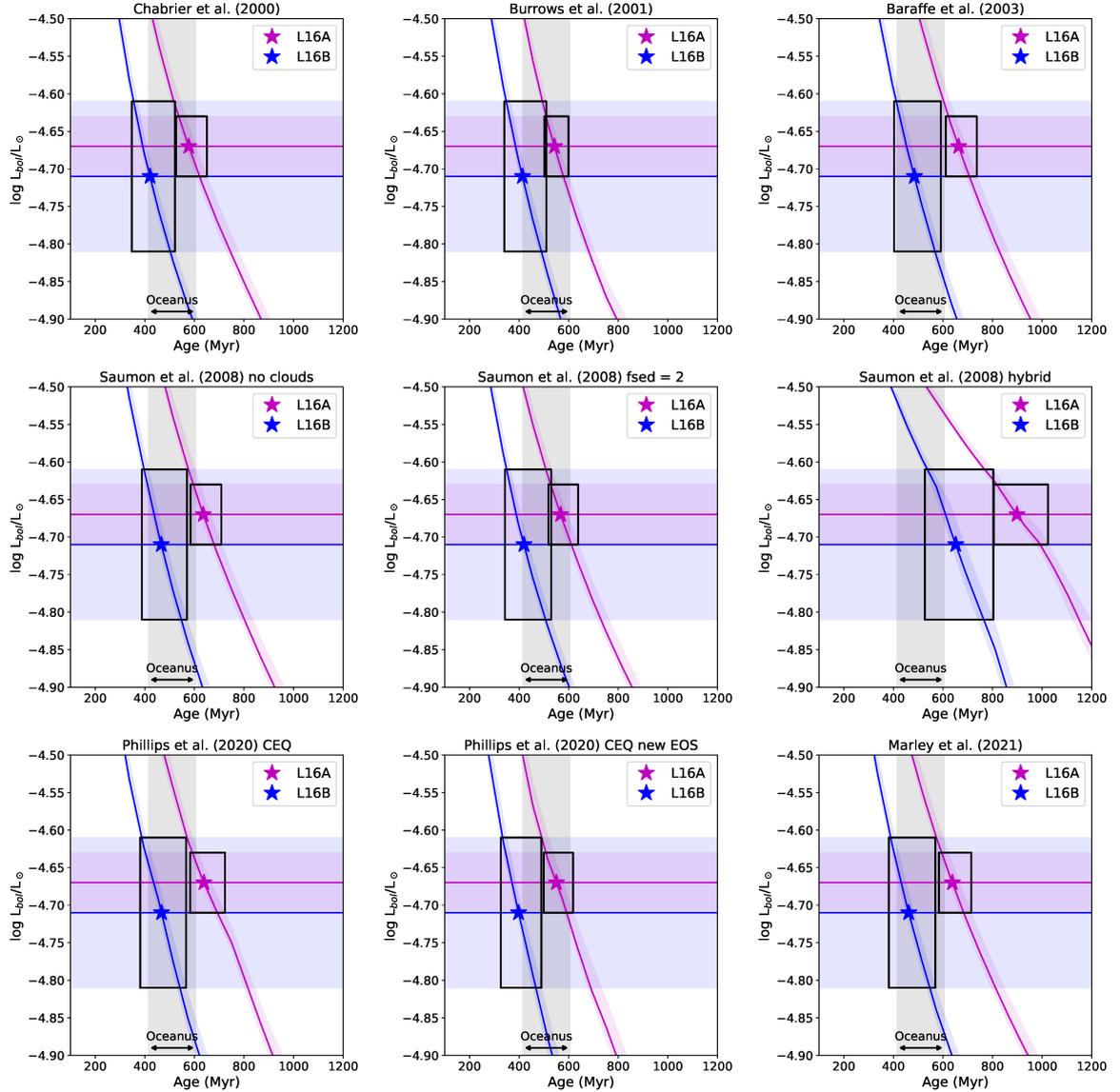

**FIGURE 8** Solar-metallicity evolutionary model predictions for the luminosities of Luhman 16 AB based on their component masses (Table 2 ) as a function of age (curved vertical lines), compared to measured component luminosity measurements (horizontal lines with shading indicating uncertainty) as reported by Faherty et al. (2014). Quantities are color-coded by primary (magenta) and secondary (blue) component. The stars indicate the intersection of models and measurements, while the black outline boxes encompass the uncertainty ranges of the measurements. We also highlight in each panel the estimated age of the Oceanus Moving Group as reported by Gagné et al. (2023, vertical grey band). The models shown from left to right are: (top panel): Chabrier et al. (2000), Burrows et al. (2001), and Baraffe et al. (2003); (middle panel): Saumon and Marley (2008) for no clouds, condensate rainout with sedimentation efficiency $f_{sed} = 2$, and "hybrid" cloud model; and (bottom panel): Phillips et al. (2020) assuming chemical equilibrium (CEQ) with original and updated equation of state by Chabrier et al. (2019), and Marley et al. (2021).

3. We used a numerical integration scheme in a three-body system to explore the possibility that prior astrometric residuals are caused by a low-mass third body. We firmly exclude the presence of any third body revolving around either component with orbital periods of 2–5000 days and masses $\gtrsim 50\,M_\oplus$ (i.e., about $3\,M_\Psi$). We can also exclude planets with masses above $1.5\,M_\Psi$ and periods greater than 400 days, however, we cannot exclude the presence of a third body with orbital periods of 2–400 days and masses $\lesssim 3\,M_\Psi$.

4. We used our accurate component masses and prior luminosity measurements to conduct a coevality test of several brown dwarf evolutionary model sets. Like previous studies, we find slight disagreement between the ages of the two components across all of the models, although the typical age ranges (400–800 Myr) are consistent with the proposed membership of Luhman 16 AB in the newly-identified 510±95 Myr Oceanus Moving Group (Gagné et al., 2023).

We finally note that the orbital fit is still imperfect, showing larger residuals at the beginning and at the end of the observed arc. Future observations, with more data points after the inversion of the orbital motion (after ∼2020), will further refine the orbital solution.

## 9 | ACKNOWLEDGMENTS

We dedicate this paper to the memory of our colleague Dr. Dimitri Pourbaix (★ 22 May 1969 — † 14 November 2021), a highly accomplished astronomer and expert in binary stars studies, who passed away during this project.

We warmly thank Shelly Meyett and Peter McCullough at STScI, our Program Coordinator and Contact Scientist for their great support during the complicated planning of the multiyear observations.

LRB and MG acknowledge support by MIUR under PRIN programme #2017Z2HSMF and by PRIN-INAF 2019 under programme #10-Bedin. The results reported herein benefited from collaborations and/or information exchange within the program "Alien Earths" (supported by the National Aeronautics and Space Administration under Agreement No. 80NSSC21K0593) for NASA's Nexus for Exoplanet System Science (NExSS) research coordination network sponsored by NASA's Science Mission Directorate. JD, DA and AB acknowledge support from STScI grants GO-13748, GO-14330 and GO-15884.

We thank an anonymous referee for the careful reading and the comments on our manuscript.

This work has made use of data from the European Space Agency (ESA) mission *Gaia* (https://www.cosmos.esa.int/gaia), processed by the *Gaia* Data Processing and Analysis Consortium (DPAC, https://www.cosmos.esa.int/web/gaia/dpac/consortium). Funding for the DPAC has been provided by national institutions, in particular the institutions participating in the *Gaia* Multilateral Agreement.

## APPENDIX A: *ESO R*-BAND PLATE

When assessing the masses for a binary system it is important to complete as much as possible the orbit, as it directly connects with the measured period and semi-major axis of the orbit, therefore it is worth the effort to scrutinize all available astronomical archives of images searching for earlier epochs of the system, possibly offering a good grip to close the orbit.

The only potentially useful epoch we were able to find was the one of the 2MASS cataogue (∼1999). Unfortunately the much needed raw data of Luhman 16 in the 2MASS survey, are not available anymore, and the pipeline-reduced measurements are not optimized for two partially-blended sources as Luhman 16 AB are. The final 2MASS catalog, nor the reduced atlas, are of sufficient quality for the present investigation.

Garcia et al. (2017) were able to find a ∼30 old plate with the two BDs detected, and that was a fundamental data point in their work to remove the mass-periodicity-eccentricity degeneracy in their model to determine the orbit and the masses given the limited arc of the orbit covered by the other data points available at that time. Garcia et al. kindly provided us with the same electronic material and we were able to perform an independent reduction of the very same digitalization of the same ESO plate.

The plate was collected in March 5, 1984, at the ESO 1m Schmidt Telescope. The exposure started at MJD 45764.955 employing the RG630-filter, IIIaF-emulsion, and an exposure time of 120 s. We derived a pixel scale of 674.67 mas/pixel using *Gaia* DR2 astrometry of sources in the field. In top panel of Fig. A1 we show the portion of the ESO plate interested by the motion of Luhman 16 AB over a period of interest for the studied epochs (∼1980-2020); the panels below zoom in on the region where Luhman 16 was in 1984.

Our independent reduction of this very same digitalized frame was conducted with the astrometric algorithms developed for ground-based wide-field imagers described in Anderson, Bedin, Piotto, Yadav, and Bellini (2006). Particular care was given to the empirical model of the PSFs, as photographic plates suffer from many problems such as reciprocity and heavy deviations from linearity. Therefore, we used only stars with similar brightness (±0.25 mag) to our targets and spatially close (∼ $4'\times 4'$) to avoid spatial PSF variations. The last row of panels in Fig. A1 shows from left to right, the digitalized ESO-plate with the fitted position for the considered sources, the model of those fitted sources, the image with the model subtracted, and finally an high-angular resolution view of the region taken with *HST* WFC3/UVIS in filter F814W at the most recent epochs. Interestingly this *HST* image shows that there are no other detectable sources at the same position that would perturb the measured astrometry of Luhman 16 A and B at the epoch when that plate was taken.

The consistency in the positions of the Gaia DR3 sources, transformed back in time to the epoch of the plate following the procedures in Bedin and Fontanive (2018), with the positions we measured for sources on this plate corroborates the accuracy of our PSF model. We find a 1-$\sigma$ ≃100 mas astrometric uncertainty based on this analysis.

The extremely different colors of the Luhman 16 AB components with respect to stars in the field for which the PSF models were extracted put serious limitations on the reliability of the positions measured on this plate. While random errors are estimated at 100 mas, systematic errors for the targeted red objects could be even larger. Furthermore, our empirical PSF model shows significantly larger residuals when compared to the the shapes of the PSFs for Luhman 16 A and B. Attempts to detect systematic chromatic refraction residuals were prevented by the lack of other stars with color similar to the targets in the field, and any extrapolation of color dependency would be unwise.

For this reason, we determined that these data points were not useful to constrain the astrometric parameters. Including the 1984 plate actually worsened our astrometric solution, giving residuals of 0.8 mas in $\alpha \cos\delta$ and 3.9 mas in $\delta$, compared to equivalent residuals of 0.5 mas in both directions without the use of the plate. On one hand, this seems rather reasonable, as a 1-$\sigma$ error of 100 mas in position even when diluted over 35 yrs (1984-2019) can lead to residuals of about 3 mas per coordinate. On the other hand, relative positions could be used to constrain the period of the orbital motion given that color terms should cancel out between the targets. Indeed, our independent estimate of the A-B relative position $(\alpha \cos\delta^{BA}, \delta^{BA}) = (-1470, 1525) \pm (101, 102)$ mas is in agreement within $\sim 2\sigma$ with that measured by Garcia et al. (2017). Nevertheless, including the ESO plate significantly worsened the residuals of our orbital solution compared to that obtained with *HST*-only data points, and we chose not to use information from this plate for our orbit fit.

## APPENDIX B: *GAIA DR2*/DR3

Both components of the Luhman 16 AB system are present in *Gaia* DR2 with the given `Identifier:GaiaDR2: 5353626573555863424` and `5353626573562355584`, respectively. Somehow `5353626573562355584` is not present in *Gaia* EDR3 or *Gaia* DR3. However, in *Gaia* DR2 neither their parallaxes nor proper motions are present. Interestingly enough, not even their relative positions are meaningful. Some complex interplay between large proper motions, large parallaxes, relatively similar magnitudes (just brighter than $G \sim 17$), extreme colors and the small separation, made *Gaia*'s automatic algorithms fail to provide reasonable values, at least in the publicly released DR2 catalog. Therefore, in the present work, we are not using in any way the positions and magnitudes in the *Gaia* DR2 for the Luhman 16 AB components.

## REFERENCES

Anderson, J. (2022, July), One-Pass HST Photometry with hst1pass., Instrument Science Report ACS 2022-02.

Anderson, J., & Bedin, L. R. (2010, September), *PASP*, *122*(895), 1035. doi:

Anderson, J., & Bedin, L. R. (2017, September), *MNRAS*, *470*(1), 948-963. doi:

Anderson, J., Bedin, L. R., Piotto, G., Yadav, R. S., & Bellini, A. (2006, August), *A&A*, *454*(3), 1029-1045. doi:

Apai, D., Nardiello, D., & Bedin, L. R. (2021, January), *ApJ*, *906*(1), 64. doi:

Baraffe, I., Chabrier, G., Allard, F., & Hauschildt, P. (2001), Pre-Main Sequence Models for Low-Mass Stars and Brown Dwarfs. In T. Montmerle & P. André (Eds.), From Darkness to Light: Origin and Evolution of Young Stellar Clusters Vol. 243, p. 571-+.

Baraffe, I., Chabrier, G., Barman, T. S., Allard, F., & Hauschildt, P. H. (2003, May), *A&A*, *402*, 701-712. doi:

Barron, E. G., Kaplan, G. H., Bangert, J., Bartlett, J. L., Puatua, W., Harris, W., & Barrett, P. (2011, January), Naval Observatory Vector Astrometry Software (NOVAS) Version 3.1, Introducing a Python Edition. In American Astronomical Society Meeting Abstracts #217 Vol. 217, p. 344.14.

Bedin, L. R., & Fontanive, C. (2018, December), *MNRAS*, *481*(4), 5339-5349. doi:

Bedin, L. R., & Fontanive, C. (2020, May), *MNRAS*, *494*(2), 2068-2075. doi:

Bedin, L. R., Pourbaix, D., Apai, D., Burgasser, A. J., Buenzli, E., Boffin, H. M. J., & Libralato, M. (2017, September), *MNRAS*, *470*(1), 1140-1155. doi:

Bellini, A., Anderson, J., & Bedin, L. R. (2011, May), *PASP*, *123*(903), 622. doi:

Bellini, A., & Bedin, L. R. (2009, December), *PASP*, *121*(886), 1419. doi:

Boffin, H. M. J., Pourbaix, D., Mužić, K. et al. (2014, January), *A&A*, *561*, L4. doi:

Branch, M. A., Coleman, T. F., & Li, V. (1999), *SIAM Journal on Scientific Computing*, *21*, 1-23.

Burgasser, A. J., Sheppard, S. S., & Luhman, K. L. (2013, August), *ApJ*, *772*(2), 129. doi:

Burgasser, A. J., & Splat Development Team. (2017), The SpeX Prism Library Analysis Toolkit (SPLAT): A Data Curation Model. In Astronomical Society of India Conference Series Vol. 14, p. 7-12.

Burrows, A., Hubbard, W. B., Lunine, J. I., & Liebert, J. (2001, July), *Reviews of Modern Physics*, *73*, 719-765.

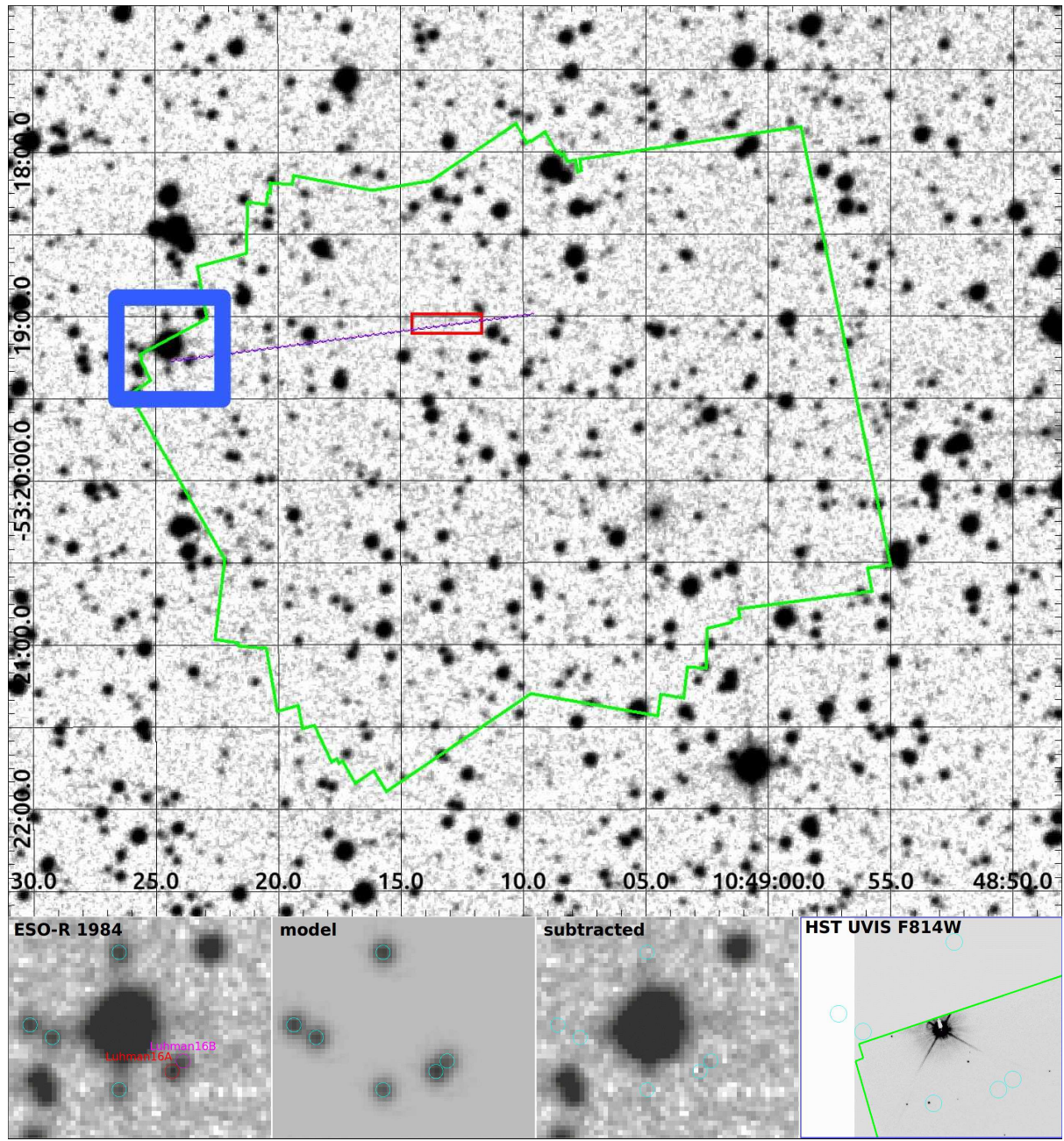

**FIGURE A1** *(Top:)* a portion of the ESO plate collected in the 1984, containing the entire region monitored by *HST* (shown in green). The rectangular in red is the region shown in Fig. 1. The region around Luhman 16 AB at this epoch is highlighted in blue (of approximately 40×30 arcsec). *(Bottom from Left to Right:)* the region around Luhman 16 AB as seen in ESO's 1984 plate (ESO-R 1984), modelling the sources in green with comparable luminosity to LUH16 (model), after the subtraction of the modelling (subtracted), and as the same region appear in *HST* F814W staked image (HST UVIS F814W).

Chabrier, G., Baraffe, I., Allard, F., & Hauschildt, P. (2000, October), *ApJ*, *542*, 464-472. doi:

Chabrier, G., Mazevet, S., & Soubiran, F. (2019, February), *ApJ*, *872*(1), 51. doi:

Chambers, J. E. (1999, April), *MNRAS*, *304*(4), 793-799. doi:

Crossfield, I. J. M., Biller, B., Schlieder, J. E. et al. (2014, January), *Nature*, *505*, 654-656. doi:

Doyle, L. R., Carter, J. A., Fabrycky, D. C. et al. (2011, September), *Science*, *333*(6049), 1602. doi:

Dupuy, T. J., & Liu, M. C. (2017, August), *ApJS*, *231*(2), 15. doi:

Eggleton, P., & Kiseleva, L. (1995, December), *ApJ*, *455*, 640-+. doi:

Faherty, J. K., Beletsky, Y., Burgasser, A. J., Tinney, C., Osip, D. J., Filippazzo, J. C., & Simcoe, R. A. (2014, August), *ApJ*, *790*, 90. doi:

Fontanive, C., Bedin, L. R., & Bardalez Gagliuffi, D. C. (2021,

February), *MNRAS*, *501*(1), 911-915. doi:

Foreman-Mackey, D., Hogg, D. W., Lang, D., & Goodman, J. (2013, March), *PASP*, *125*(925), 306. doi:

Foreman-Mackey, D. (2016, jun), *The Journal of Open Source Software*, *1*(2), 24. Retrieved from https://doi.org/10.21105/joss.00024 doi:

Gagné, J., Moranta, L., Faherty, J. K. et al. (2023, March), *ApJ*, *945*(2), 119. doi:

Gaia Collaboration, Vallenari, A., Brown, A. G. A. et al. (2023, June), *A&A*, *674*, A1. doi:

Garcia, E. V., Ammons, S. M., Salama, M. et al. (2017, September), *ApJ*, *846*(2), 97. doi:

Gillon, M., Triaud, A. H. M. J., Jehin, E. et al. (2013, July), *A&A*, *555*, L5. doi:

Heintz, W. D. 1978, Double stars (Vol. 15).

Holman, M. J., & Wiegert, P. A. (1999, January), *AJ*, *117*(1), 621-628. doi:

Kaplan, G., Bartlett, J., Monet, A., Bangert, J., & Puatua, J. (2011, March), *User's Guide to NOVAS Version F3.1.*, Washington, DC: USNO.

Karalidi, T., Apai, D., Marley, M. S., & Buenzli, E. (2016, July), *ApJ*, *825*(2), 90. doi:

Kirkpatrick, J. D., Cushing, M. C., Gelino, C. R. et al. (2011, December), *ApJS*, *197*, 19. doi:

Kniazev, A. Y., Vaisanen, P., Mužić, K. et al. (2013, June), *ApJ*, *770*(2), 124. doi:

Lindegren, L., Bastian, U., Biermann, M. et al. (2021, May), *A&A*, *649*, A4. doi:

Liu, M. C., Dupuy, T. J., & Leggett, S. K. (2010, October), *ApJ*, *722*, 311-328. doi:

Lodieu, N., Zapatero Osorio, M. R., Rebolo, R., Béjar, V. J. S., Pavlenko, Y., & Pérez-Garrido, A. (2015, September), *A&A*, *581*, A73. doi:

Luhman, K. L. (2013, April), *ApJ*, *767*(1), L1. doi:

Marley, M. S., Saumon, D., Visscher, C. et al. (2021, October), *ApJ*, *920*(2), 85. doi:

Moré, J. J., Garbow, B. S., & Hillstrom, K. E. (1980), *User Guide for MINPACK-1.*, Argonne National Laboratory Report ANL-80-74, Argonne, Ill., 1980.

Phillips, M. W., Tremblin, P., Baraffe, I. et al. (2020, May), *A&A*, *637*, A38. doi:

Pourbaix, D. (1994, October), *A&A*, *290*, 682-691.

Pourbaix, D. (1998a), on *Orbit determination of binary stars using simulated annealing*. In: De Leone, R., Murli, A., Pardalos, P.M., Toraldo, G. (eds.), High Performance Algorithms and Software in Nonlinear Optimization. Applied OptimizationWashington, vol 24. Springer, Boston, MA.

Pourbaix, D. (1998b, August), *A&AS*, *131*, 377-382. doi:

Pourbaix, D. (2000), in 'Encyclopedia of optimization', Eds. C.A. Floudas and P.M. Pardalos, Kluwer Academic Publishers.

Rein, H., Hernandez, D. M., Tamayo, D. et al. (2019, June), *MNRAS*, *485*(4), 5490-5497. doi:

Rein, H., & Liu, S. F. (2012, January), *A&A*, *537*, A128. doi:

Rein, H., & Spiegel, D. S. (2015, January), *MNRAS*, *446*(2), 1424-1437. doi:

Rein, H., & Tamayo, D. (2015, September), *MNRAS*, *452*(1), 376-388. doi:

Riess, A. G., Casertano, S., Anderson, J., MacKenty, J., & Filippenko, A. V. (2014, April), *ApJ*, *785*(2), 161. doi:

Sahlmann, J., & Lazorenko, P. F. (2015, October), *MNRAS*, *453*(1), L103-L107. doi:

Saumon, D., & Marley, M. S. (2008, December), *ApJ*, *689*, 1327-1344. doi:

Wright, E. L., Eisenhardt, P. R. M., Mainzer, A. K. et al. (2010, December), *AJ*, *140*, 1868-1881. doi:

Zechmeister, M., & Kürster, M. (2009, March), *A&A*, *496*(2), 577-584. doi: